\documentclass[onecolumn]{aastex61}

\received{July 1, 2016}
\revised{September 27, 2016}
\accepted{\today}
\shorttitle{Supercritical Accretion onto Magnetized Neutron Star}
\shortauthors{H. R. Takahashi \& Ken Ohsuga}
\begin{document}

\title{General Relativistic Radiation Magnetohydrodynamics
Simulations of Supercritical Accretion onto Magnetized Neutron Star;
-Modelling of ultra luminous X-ray pulsars}
\email{takahashi@cfca.jp}

\author[0000-0003-0114-5378]{Hiroyuki R. Takahashi}
\affil{Center for Computational Astrophysics, National
  Astronomical Observatory of Japan, National Institutes of Natural
  Sciences, Mitaka, Tokyo 181-8588, Japan}

\author{Ken Ohsuga}
\affil{Division of Theoretical Astronomy, National
  Astronomical Observatory of Japan,
  National Institutes of Natural
  Sciences, Mitaka, Tokyo 181-8588, Japan}
  \affil{School of Physical Sciences,Graduate University of
Advanced Study (SOKENDAI), Shonan Village, Hayama, Kanagawa 240-0193,
Japan}



\begin{abstract}
By performing 2.5-dimensional general relativistic radiation
 magnetohydrodynamic simulations,
 we demonstrate supercritical accretion onto a non-rotating, magnetized
 neutron star,
 where the magnetic field strength of dipole fields
 is $10^{10}\ \mathrm{G}$ on the star surface.
 We found the supercritical accretion flow consists of two parts;
 the accretion columns and the truncated accretion disk.
 The supercritical accretion disk,
 which appears far from the neutron star,
 is truncated at around $\simeq 3R_\mathrm{*}$
 ($R_*=10^6 \mathrm{cm}$ is the neutron star radius),
 where the magnetic pressure via the dipole magnetic fields
 balances with the radiation pressure of the disks.
 The angular momentum of the disk around the truncation
 radius is effectively transported inward 
 through magnetic torque by dipole fields,
 inducing the spin up of a neutron star.
 The evaluated spin up rate,
 $\sim -10^{-11}\, {\rm s}\,{\rm s}^{-1}$,
 is consistent with the recent observations 
 of the ultra luminous X-ray pulsars. 
 Within the truncation radius,
 the gas falls onto neutron star along dipole fields,
 which results in a formation of accretion columns onto
 north and south hemispheres.
 The net accretion rate
 and the luminosity of the column
 are $\simeq 66L_{\rm Edd}/c^2$ and $\lesssim 10L_{\rm Edd}$,
 where $L_{\rm Edd}$ is the Eddington luminosity and
 $c$ is the light speed.
 Our simulations support a hypothesis whereby
 the ultra luminous X-ray pulsars are powered by
 the supercritical accretion onto the 
 magnetized neutron stars.
\end{abstract}




\section{Introduction} \label{sec:intro}
Accretion disks around compact objects such as neutron stars or black
holes are one of the most energetic system in the universe. The disk
is powered by liberating the gravitational energy, so that the mass
accretion rate is an important key parameter determining their activities. 
Theoretically, there are three distinct accretion modes according to mass
accretion rate, e.g., the
radiatively inefficient accretion flow \citep{1994ApJ...428L..13N},
standard disk \citep{1976MNRAS.175..613S}, and slim disk
\citep{1988ApJ...332..646A}. 
The mass accretion rate of the slim disk,
which is the highest among three modes,
is higher than the critical rate ($\dot{M}_{\rm crit}=L_{\rm Edd}/c^2$) 
where $L_\mathrm{Edd}$ is the Eddington luminosity
and $c$ is the light speed.
The luminosity of the slim disk 
exceeds the Eddington luminosity,
so that the slim disk is applied to 
the very luminous objects.
One of the possible site of the
supercritical accretion is 
the Ultra luminous X-ray sources (ULXs).
It is well known that ULXs show
super-Eddington luminosity for the stellar mass black hole.
Thus, it has been suggested that 
the supercritical accretion occurs in the ULXs.
But the another solution with sub-critical accretion onto 
an intermediate mass black hole is possible. 
It is still under debate 
which is feasible to explain characteristics of ULXs.

A recent finding of pulsation in ULXs changes the situation. 
\cite{2014Natur.514..202B} found a X-ray pulsation with
period of $P=1.37\ \mathrm{s}$ and its time derivative $\dot P=-2\times
10^{-10}\ \mathrm{s\ s^{-1}}$ from M82 X-2. After this notable discovery, X-ray
pulsations have been observed in other two ULXs,
of which the spin up rate is reported to be 
$\sim -4 \times 10^{-11}\, {\rm s}\,{\rm s}^{-1}$
and $\sim -5 \times 10^{-9}\, {\rm s}\,{\rm s}^{-1}$
\citep{2016arXiv160907375I, 2016ApJ...831L..14F, 2017MNRAS.466L..48I}.
These observational facts indicate that a supercritical accretion 
onto a pulsar (namely, ULX Pulsars), since the luminosity of these ULXs
highly exceeds the Eddington luminosity for the neutron stars.


\cite{2007PASJ...59.1033O}
performed radiation hydrodynamics simulations 
of supercritical disk accretion onto 
a non-magnetized neutron star. 
After that \cite{Takahashi17}
performed general relativistic radiation magnetohydrodynamics 
(GR-RMHD) simulations of the supercritical disks
around a non-magnetized neutron star. 
They showed a formation of powerful outflows from 
the vicinity of the neutron star. 
Since the neutron star, different to the
black hole, does not swallow the mass and energy,
the resulting outflow power is stronger than the black hole 
\citep[see, also][]{2016MNRAS.458L..10K}. 
However, in these study, the neutron star magnetic field is ignored,
although observations of the $P-\dot P$ relations
reveal the existence of the strong field
\citep{2014Natur.514..202B}.

If the neutron star magnetic field
is strong enough to prevent the disk accretion,
the matter would fall onto the neutron star
along the magnetic field lines.
\cite{1976MNRAS.175..395B} proposed a model of 
the supercritical column accretion onto the neutron star.
Also, \cite{2016PASJ...68...83K} performed 
radiation hydrodynamic simulation 
of supercritical accretion column.
The interaction between neutron star magnetic fields 
and accretion disks is, however, unresolved in
such local simulations. 
Thus, the global radiation magnetohydrodynamic simulation is necessary to 
reveal the global structure of the supercritical accretion flows around
the neutron star by taking into consideration
the interaction between radiation dominated accretion disks and neutron
star magnetosphere. 
Then, the general relativistic effects also should be taken into account,
since the neutron star radius is comparable to its gravitational radius,
and since Alfv\'en speed in the magnetosphere would be close 
to the light speed via the strong magnetic field.
Fortunately, GR-RMHD code has been developed by authors
\citep{2014MNRAS.441.3177M, 2014MNRAS.439..503S, 2015MNRAS.454.2372S,
2016ApJ...826..23, 2017MNRAS.466..705S}. 
The previous work, including this study, assume the equation of
state for radiation field to close the system
\citep{1984JQSRT..31..149L}.
The M-1 closure is useful to include effects of radiation
such as energy and momentum exchange between the gas and radiation,
but there are some limitations in this approximation. The discussion on
this limitation is shown in section 4.

In this paper, we adopted the moment formalism on radiation field
for the first step and we report our results of 2.5-dimensional global
GR-RMHD simulations of the supercritical accretion flows around the
neutron stars with strong dipole magnetic field.
Our results show that the supercritical accretion onto the
magnetized neutron star is plausible model
for the ULX pulsars.

\section{Methods}
We solved GR-RMHD equations by adopting moment formalism describing
radiation field \citep{1981MNRAS.194..439T}. Basic equations and
numerical schemes are the same with our former papers
\citep{2016ApJ...826..23}. We 
incorporate the mean electron scattering, free-free emission, thermal
Comptonization \citep{2015MNRAS.447...49S}, and synchrotron emission
\citep{2017MNRAS.466..705S} for the source of opacity.

We solved GR-RMHD equations in polar coordinate $(t,r,\theta,\phi)$ with
Boyer-Lindquist metric by assuming axisymmetry with respect to a
rotation axis $\theta=0, \pi$. The computational box consists of
$r=[R_*, 199R_*]$ and $\theta=[0,\pi]$, where $R_* = 10\ \mathrm{km}$
is a neutron star radius. 
The radial grid exponentially increases with
radius and the uniform grid is adopted in $\theta-$direction. 
Numerical grid points are $(N_r, N_\theta, N_\phi)=(528, 288, 1)$.

The neutron star mass is assumed to be $M_*=1.4M_\sun$.
We adopt a dipole magnetic field \citep{1983ApJ...265.1036W},
of which strength is fixed to be $B_*=10^{10} \mathrm{G}$
at the neutron star surface.
The magnetic axis is assumed to coincide with 
the rotation axis.
We suppose the non-rotating neutron star in the present study
for simplicity.
This assumption would be valid
since the observed rotation velocity of the neutron star 
is slower than the Keplerian velocity.
The study for the case of the rapidly rotating neutron star
is left as important future work.

We set an equilibrium torus given by \cite{1976ApJ...207..962F}, but the
gas pressure is replaced by the sum of gas $(p_\mathrm{gas})$ and radiation
$(p_\mathrm{rad})$ pressure
assuming local thermodynamic equilibrium. 
The inner edge and pressure maximum of the torus are located at
$r=10R_*$ and $r=15R_*$, respectively. 
We employ the maximum density of the torus $\rho_0$
of $0.1 \mathrm{g\ cm^{-3}}$.
In addition to the dipole magnetic field of the neutron star,
the poloidal magnetic field
confined in this torus is embedded so that its magnetic flux vector
$A_\phi$ is proportional to the density. 
The confined magnetic fields are anti-parallel to
dipole fields at the inner part of the torus
\citep[but, see][for the difference of parallel
and anti-parallel cases]{2011MNRAS.416..416R}. 
The ratio of maximum
$p_\mathrm{gas}+p_\mathrm{rad}$ to the maximum magnetic pressure
($p_\mathrm{mag}$) via the confined magnetic fields 
is set to be 100. 
We also give a perturbation on $p_\mathrm{gas}+p_\mathrm{rad}$ by its
10\% to break an equilibrium state.

The neutron star and the torus are initially surrounded by 
a relatively low density corona.
Its density decreases with radius as $\rho = \rho_c (R_*/r)^3$,
where $\rho_c$ is set as an initial coronal 
$\sigma$ parameter becomes $\sigma_c=B_*^2/(4\pi \rho_c c^2)=500$.
The gas pressure of the corona is $p_\mathrm{gas} = 0.1 \rho c^2$. 


We adopted 
a symmetric boundary condition at $\theta = 0, \pi$, and
outflow boundary condition at $r=199R_*$. 
For the inner boundary ($r=R_*$), 
the velocity and radiation flux are set to be zero.
A free boundary condition is applied to the other quantities
\citep{2011MNRAS.416..416R}.
The outgoing boundary condition is applied at outer boundary.

\section{Results}
\begin{figure*}
 \begin{center}
 \includegraphics[width=18cm]{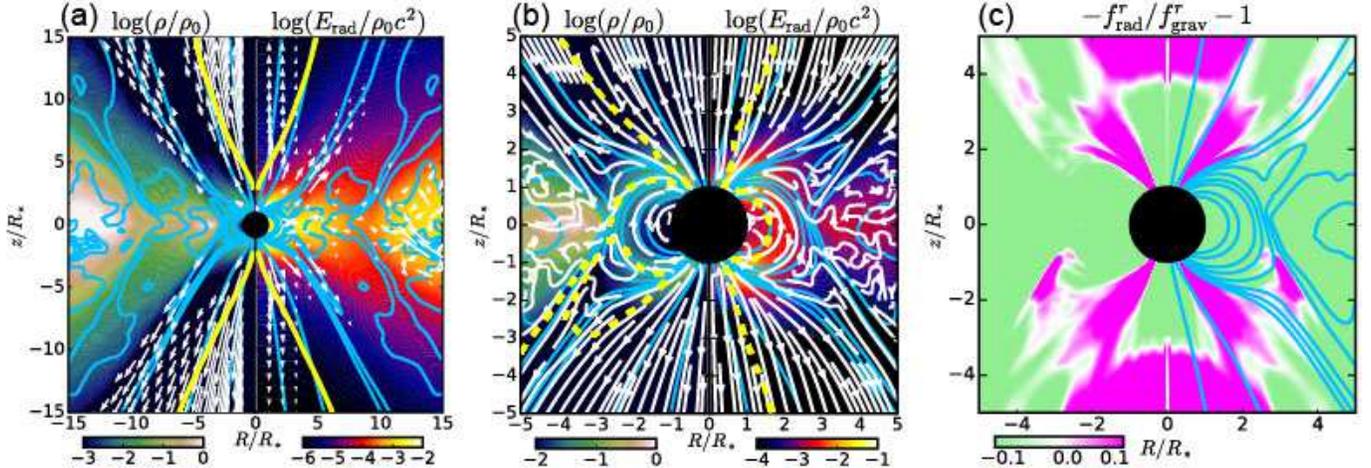}
  \caption{a: Colors show $\rho/\rho_0$ (left) and
 $E_\mathrm{rad}/\rho_0c^2$ (right), while vectors show the four
  velocity (left) and and radiation flux (right). Blue lines are magnetic field
  lines and yellow lines show the photosphere measured from poles.
  b: enlarged view of figure 1a, but white curves show stream lines of four
  velocity (left) and radiation flux (right). The yellow dashed curves
  show $\beta=1$ (left) and $\sigma=1$ (right). 
  c: Color shows $-f_\mathrm{rad}^r/f_\mathrm{grav}^r - 1$.} 
\label{fig1}
 \end{center}
\end{figure*}

After simulation starts, an initial equilibrium state gradually breaks
due to a given perturbation. The accreting matter reaches 
a neutron star surface at $t=4,800t_\mathrm{g}$, 
where $t_\mathrm{g}=GM_*/c^3=6.9\times 10^{-6}\
\mathrm{s}$ is the light crossing time of the
gravitational radius. 
After that, the mass accretion rate is almost kept constant
until simulation stops at $t=15,000t_\mathrm{g}$.
Figure 1a shows global structure of 
inflow and outflow around magnetized neutron star
on $(R,z)=(r\sin\theta, r\cos\theta)$ plane.
Color shows $\rho/\rho_0$ (left) and $E_\mathrm{rad}/\rho_0 c^2$ (right), while
arrows show four velocity of fluid (left) and radiation flux (right). 
Blue lines indicate magnetic field lines.
In figures 1, 2, and table 1, all the data 
are averaged in time between 
$[10^4 t_\mathrm{g}, 1.5 \times 10^4 t_\mathrm{g}]$.
We find an accretion disk is formed at the region of 
$3R_* \lesssim r \lesssim 8R_*$ (white region 
near the equatorial plane). 
The disk density on the equatorial plane is about 
$0.1\rho_0=0.01\ \mathrm{g\ cm^{-3}}$.
The radiation energy is enhanced in the disk
(see yellow region in the right panel) and
dominates the other energy.
Inside the disk, magnetic fields are tangled due to the
development of magnetorotational instability (MRI)
\citep{1991ApJ...376..223H,1991ApJ...376..214B,1998RvMP...70....1B}.
The induced turbulent motion is clearly shown in 
the left panel in figure 1b (stream lines of four velocity shown with
white curves).
Here we note that the high density region at $r \gtrsim 8R_*$
is the remnant of the initial torus.

White arrows in the left panel 
means that strong jets are ejected around the rotation axis. 
The jets have an opening angle of $\lesssim 18^\circ$ and
typical speed is $0.4c$ at $|z|\sim 30R_*$. 
These fast jets are ejected
due to the radiation force \citep{2015PASJ...67...60T}. 
Indeed, we find that the outward radiation flux
is enhanced around the rotation axis.
Besides fast jets, relatively slow and dense outflows 
are ejected in the direction of $\theta = 40^\circ$ and $140^\circ$, 
even though
the boundary between the jet and the outflow is not clear 
since the density as well as the velocity smoothly change in theta-directions.
Such outflows have velocity of $\simeq 0.1c$ around $r=30 R_*$, and its
density is about two orders of magnitude higher than fast jets. 
The outflows are also driven by the radiation force.
We find the slightly enhanced radiation flux 
in the outflow region (see right panel).
In outflow region, the recurrent magnetic reconnections occur 
\citep{2006A&A...453..785F, 2011MNRAS.416..416R} 
even though we applied ideal MHD.
The reconnections also help the outflow acceleration
\citep{1996ApJ...468L..37H, 2004ApJ...600..338K}.

\begin{figure}
 \begin{center}
 \includegraphics[width=15cm]{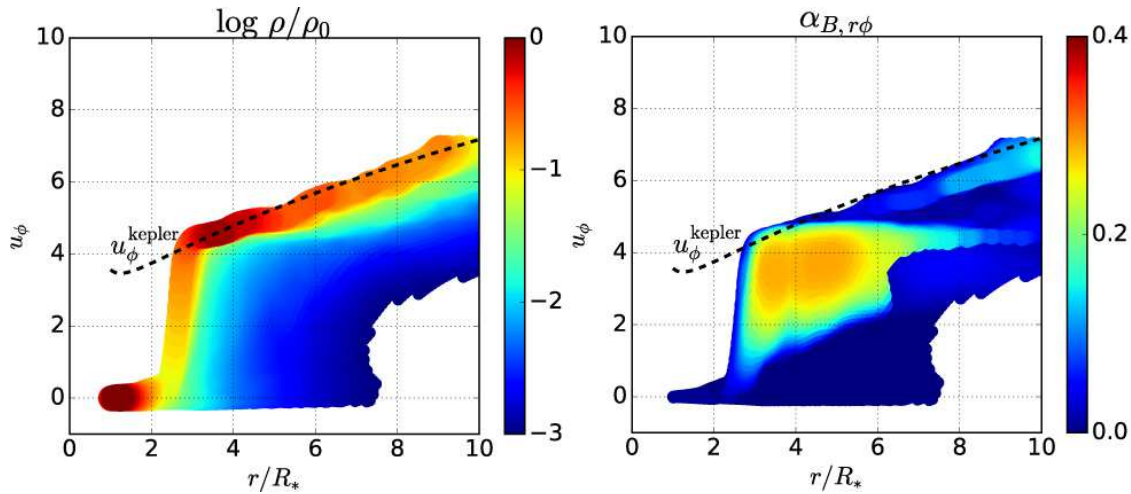}
  \caption{Density (left) and $\alpha_{\mathrm{B},r\phi}$ (right) plots on
  $r-u_\phi$ plane. Black dashed line shows Keplerian velocity.} 
\label{fig2}
 \end{center}
\end{figure}

For a smaller radius, an interaction of dipole magnetic field and
disk is manifest. 
 Figure 1b is an enlarged view of figure 1a.
We plot by yellow dashed lines
$\beta = (p_\mathrm{gas}+p_\mathrm{rad})/p_\mathrm{mag}=1$
(left panel) and $\sigma\equiv
b^2/(4\pi\rho c^2)=1$, where $b^2=8 \pi p_\mathrm{mag}$
(right panel).
The magnetic energy dominates the other energy 
around the rotation axis.
In addition, we find the region of $\beta<1$ near the neutron star 
for $25^\circ \lesssim \theta \lesssim 155^\circ$
(magnetosphere).
The right panel in figure 1b clearly shows that 
$\sigma$ value is very large near the neutron star
as well as around the rotation axis.
Typically, we find $\sigma \sim 100$ 
deep inside the magnetosphere and 
$\sigma \sim 150$ around poles.
Since the ideal MHD is assumed in the present simulations, 
the disk matter cannot enter the magnetosphere.
Hence,
the accretion disk is truncated around $r_\mathrm{T}\simeq 3R_*$. 
At the truncation radius, the magnetic reconnection takes place
between the dipole field and disk field. As noted before, we assume that
the magnetic polarity between these fields is anti-parallel. 
The magnetic reconnection at the truncation radius leads to the mass
transfer from disks to the magnetosphere \citep{2011MNRAS.416..416R}.
Inside the truncation radius, 
the matter accretes along the magnetic field lines,
producing the accretion columns in north and south hemispheres 
around $\theta = 25^\circ$ and $155^\circ$ \citep{1979ApJ...234..296G} 
(see green region near the neutron star surface in the left panel).
The flow is turbulent in the accretion disk, while it is relatively laminar
in accretion columns (see left panel in figure 1b).
In classical theories applied for T-tauri stars, white dwarfs and
neutron stars, the magnetospheric radius is determined by a
balance between the magnetic pressure of dipole fields and the gas (plus
ram) pressure. For the supercritical flows, the
radiation pressure dominates the other energy. Thus the magnetospheric
radius of accretion disks is determined by the magnetic pressure and
the radiation pressure.

\cite{2014arXiv1410.8745L} showed that the radiation force prohibits
from mass accretion for relatively low mass accretion rate case. In
contrast, the radiation force does not prevent the mass accretion in the
accretion columns and the accretion rate largely exceeds the critical
rate in our model. Figure 1-c shows a ratio of radial component of
radiation force and gravity force. The outward radiation force is
smaller than the gravity inside the accretion columns (see green region
between $40^\circ \lesssim \theta \lesssim 140^\circ$). 
In contrast, we found the radiation force overcomes the
gravity force just above (below) the accretion column in northern
(southern) hemisphere (see magenta region between 
$10^\circ\lesssim \theta \lesssim 40^\circ$ and 
$140^\circ\lesssim \theta \lesssim 170^\circ$). This is thought to
be cased by that the radiation energy escapes from side walls of
accretion columns, which reduces (enhances) the radiation energy inside
(outside) the accretion columns. Thus, the mass accretion is feasible
inside the column, in contrast, the gas is blown away by the strong
radiation force outside the column. In addition, the large optical
thickness of the column helps the matter accretes. The optical depth
inside accretion columns largely exceeds unity, so that the radiation
energy is transported by the radiative diffusion. Then, radiation flux
and the radiation flux force tend to be suppressed. Such situation is
basically same as that in \cite{2016PASJ...68...83K} in which the
super-Eddington accretion column of which the optical depth is much
larger than unity was realized by the escape of the radiation energy
from the side wall of the column. \cite{2007ApJ...670.1283O} also reported
that the super-Eddington disk accretion is feasible via the anisotropy
and the large optical thickness.
\begin{figure}
 \begin{center}
 \includegraphics[width=10cm]{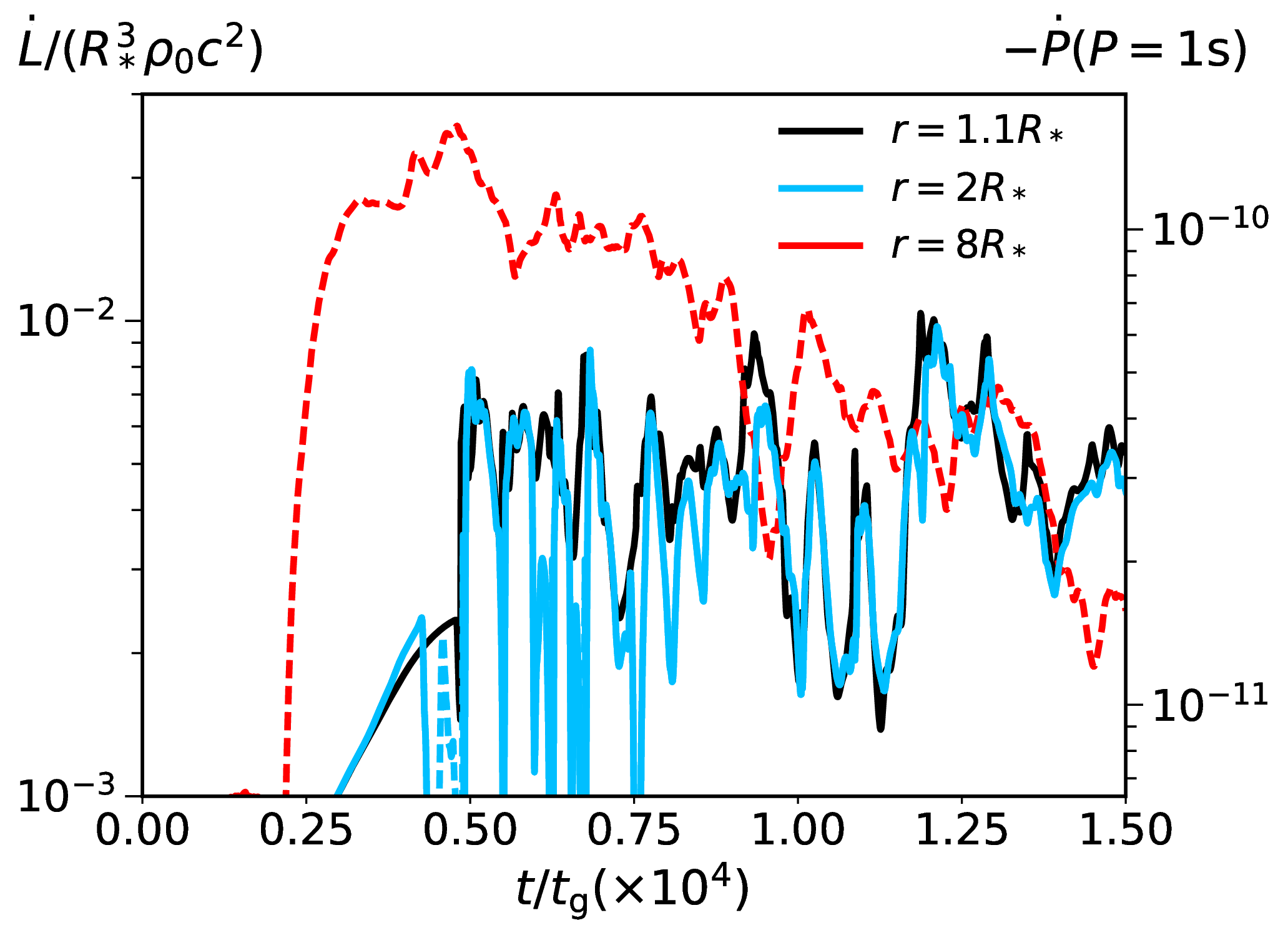}
  \caption{time evolution of angular momentum flux at $r=1.1R_*$
  (black), $r=2R_*$ (blue), $r=3R_*$ (orange) and $r=8R_*$ (red). 
  Solid lines show negative flux and dashed ones do positive flux.
} 
\label{fig3}
 \end{center}
\end{figure}


In table 1, we show mass accretion rate, outflow rate, 
net inflow rate (accretion rate - outflow rate), and
bolometric luminosity measured at $r=1.1R_*,\ 3R_*$ and $8R_*$.
%
The mass accretion rate as well as the 
net inflow rate exceeds the critical rate.
This means that 
the supercritical accretion occurs not only 
the disk ($\gtrsim 3R_*$) but also accretion columns
($\lesssim 3R_*$). 
Although the net inflow rate is around $10 \dot{M}_{\rm crit}$
at $r=1.1R_*,\ 3R_*$, it is $\sim 300 \dot{M}_{\rm crit}$ at $r=8R_*$.
This implies that 
the large amount of the accreting matter accumulates
in the disk region, $3R_* \lesssim r \lesssim 8R_*$,
and a part of the matter accretes onto the neutron star
through the accretion columns.
When the density around the truncation radius
drastically increases via the accumulation of the matter,
the flow structure might change.
For instance, the high density matter flows into the columns
and the mass accretion rate onto the star surface 
might greatly increase.
Or, the magnetosphere might be crushed by 
the strengthened gas and radiation pressure.
We need long-term simulations
in order to make clear the point.

The resulting bolometric luminosity largely exceeds the Eddington
luminosity as shown in the table 1.
Since $L_{\rm bol}$ at $r=3R_*$ is comparable to or slightly
larger than that at $r=8R_*$, 
it is thought that the radiation is mainly generated near the neutron star.
The right panel in figure 1b shows
that the radiation energy is drastically enhanced
in the accretion columns,
and a strong outward radiation flux appears
outside the columns.
The formation of very luminous accretion columns 
has also been proposed by authors
\citep{1976MNRAS.175..395B,2016PASJ...68...83K}.
At the outer region of $r\gtrsim 10R_*$, 
the outward radiation flux is enhanced around the rotation axis
(see the right panel of figure 1a).
Thus, this object would be identified as 
highly super-Eddington source for the face-on observer.

Left panel of figure 2 shows distribution of a gas density in $r-u_\phi$
plane. Here $u_\phi$ is the $\phi$-component of covariant four
velocity. 
In this figure, only a high density ($\rho > 10^{-3}\rho_0$) region is
plotted. For $r>r_\mathrm{T}$, a dense gas inside accretion
disks (red and orange) is located around the dotted line.
This means that the rotation velocity
of the disk matter is close to the Keplerian velocity. 
Around the truncation
radius, the angular momentum suddenly decreases and it 
goes to zero on the neutron star surface. 
That is, the matter in the accretion columns
rotates slowly.
This is because that the angular momentum of the accreting matter 
is quickly transported to the neutron star
via the magnetic torque of dipole fields.
This is understood from the right panel of figure 2,
in which we show a viscous parameter evaluated by
$\alpha_{B,r\phi} =T_\mathrm{mag}^{\tilde r
\tilde \phi}/(p_\mathrm{gas}+p_\mathrm{rad}+p_\mathrm{mag})$.
Here, $T_\mathrm{mag}^{\tilde r \tilde \phi}$ is a $r-\phi$ component 
of energy momentum tensor of electromagnetic field
\citep{2010MNRAS.408..752P},
and the tilde indicates a comoving frame quantity.
This panel shows $\alpha_{B,r\phi}$ is about or less than $0.1$
in the disk region (near the dotted line for $r\gtrsim 5R_*$).
Such a small viscosity is consistent with GR-MHD simulations
\citep{2013MNRAS.428.2255P}. 
We also find a relatively large $\alpha_{B,r\phi}$ region (orange)
near the truncation radius.
As shown in figure 2b, 
the strong magnetic field lines in this region
connect to the neutron star surface.
Since the non-rotating neutron star is employed in
the present work,
the magnetic torque of dipole fields effectively extract
the angular momentum of the matter.

We plot the angular momentum flux computed by
\begin{equation}
 \dot L = \int T_{\mathrm{mag} \phi}^r \sqrt{-g} d\theta d\phi,
\end{equation}
\citep{2009MNRAS.399.1802R}, 
where $g$ is the determinant of the metric. 
Black, blue, and red lines corresponds to the angular momentum
flux measured normalized by $R_*^3 \rho_0 c^2$
at $r=1.1 R_*,\ 2R_*$, and $8R_*$. 
Solid lines show inward flux and dashed lines indicate
outward flux. 
Here we only consider an angular momentum flux contributed from magnetic fields 
since the kinetic and radiation terms are negligible 
even in the radiation dominated disk.
Far from the truncation radius ($r=8R_*$), 
the angular momentum is transported
outward due to the MRI turbulence. 
Inside $r\simeq r_\mathrm{T}$ ($r=1.1R_*$ and $2R_*$), 
the angular momentum flux is negative
throughout simulations. 
Thus the angular momentum of the disk is extracted through
the dipole field, leading to 
spin up of a neutron star. 
If we assume the rotation period of neutron star to be $1\ \mathrm{s}$,
the spin up period estimated from our result $\dot L=0.04 R_*^3 \rho
c^2$ corresponds to $\dot P \sim -3\times 10^{-11}\, {\rm s}\,{\rm s}^{-1}$.
This result is consistent with the observations of ULX pulsars,
from $\sim-5 \times 10^{-9}\, {\rm s}\,{\rm s}^{-1}$
to $\sim -4 \times 10^{-11}\, {\rm s}\,{\rm s}^{-1}$.


\section{Conclusion \& Discussion}
In this paper,
we first demonstrate a supercritical accretion onto a 
non-rotating magnetized neutron star using 
2.5-dimensional GR-RMHD simulations. 
We show that radiation dominated accretion disks are formed far from a
neutron star, and are truncated around $r=r_\mathrm{T}=3R_*$. Inside
this radius, the accretion columns form and the matter accretes
onto north and south hemisphere of the neutron star.
The truncation radius of the radiation dominated disk
is determined by the radiation pressure 
and the magnetic pressure 
via the dipole fields of the neutron star.
Indeed, we find that the radiation pressure is 
comparable to the magnetic pressure at $r \sim r_\mathrm{T}$. 
Also we found fast jets ($\sim 0.4c$) along the rotation axis
($\lesssim 20^\circ$ and $\gtrsim 160^\circ$).
Besides fast jets, 
relatively slow ($\sim 0.1c$) and dense outflows are emanated 
with larger opening angle, $\sim 40^\circ$.
Both jets and outflows are accelerated by the radiation force. 

We show that a resulting luminosity exceeds the Eddington luminosity,
$\sim 10 L_\mathrm{Edd}$.
This radiation flux is marginally collimated around the disk and
magnetic rotation axis, $\theta =0$ and $\pi$. The $e$-folding angle of
the luminosity, which corresponds to the beaming factor $b$, is about
$0.2$ at $150 R_*$. This beaming factor is
larger than that proposed by \citep{2016MNRAS.458L..10K}.
They showed that the beaming
factor decreases with mass accretion rate $b \propto \dot M^{-2}$
\citep{2009ApJ...702..489R}. We do not
know the reason of this discrepancy, but one possibility is that the
massive outflows with a large opening angle decreases the beaming
factor due to the electron scattering. 
The dependence of the mass accretion rate on the luminosity, beaming factor and accretion mode will be reported in the subsequent paper. 

As we have mentioned above,
the spin up rate obtained by our simulations,
$\dot{P} \sim - 10^{-11}\,{\rm s}\,{\rm s}^{-1}$, is consistent
with the observations.
Such a suitable spin up rate is naturally understood
by the interaction of the neutron star magnetic fields 
and supercritical accretion disks.
The spin up rate is estimated by $\dot P = - \dot M l(r_\mathrm{m})/M_*
l_*$, \citep{1986bhwd.book.....S}, where $l$ and $l_*$ are angular
momentum of disk matter and neutron star, 
and $r_{\rm m}$ is the magnetospheric radius.
The magnetospheric radius is obtained 
by the balance between the magnetic pressure of dipole fields 
and the radiation pressure of the disk
for the case of the radiation dominated accretion disks.
Applying self-similar solutions for slim disk
\citep{1999PASJ...51..725W},
the magnetospheric radius is obtained as
$r_{\rm m}/R_* \simeq 1.7 (\alpha/0.1)^{2/7}(\dot{M}/10^2\dot{M}_{\rm crit})^{-2/7}
(B_*/10^{10} G)^{4/7} (M_*/1.4M_\odot)^{-3/7} 
(R_*/10{\rm km})^{5/7}$,
where $\alpha$ is the viscous parameter.
Note that the magnetospheric radius
is smaller than the corotation radius,
$r_\mathrm{cor}/R_*=170 
(M_*/1.4M_\sun)^{1/3} (P_*/1\ \mathrm{s})^{2/3}$
where $P_*$ is the rotation period of the neutron star. 
The magnetospheric radius is a decreasing function of $\dot M$.
The magnetosphere would disappear for a sufficiently high mass accretion
rate and the accretion disks reach to a neutron star surface \citep{2007PASJ...59.1033O}.
The spin up rate is estimated by assuming that the Keplerian angular
momentum at the magnetospheric radius is transported to neutron star
without dissipation, as
$\dot{P} \simeq -4.9 \times 10^{-11}
(\alpha/0.1)^{1/7}
(\dot{M}/10^2\dot{M}_{\rm crit})^{6/7}
(B_*/10^{10} G)^{2/7} 
(M_*/1.4M_\odot)^{2/7} 
(R_*/10{\rm km})^{-8/7} 
(P_*/1{\rm s})
\,{\rm s}\,{\rm s}^{-1}$.
This is comparatively near to the spin up rate
obtained by the present simulations,
and is consistent with observations of ULX pulsars.
Hence, we stress again that 
our present simulations reveal that
the ULX pulsars can be explained by
the supercritical accretion onto
the magnetized neutron stars.

The net inflow rate in the accretion columns is about ten times larger than
the critical rate,
and the resulting bolometric luminosity is several $\times L_{\rm Edd}$.
Since the outward radiative flux is 
sensitive to the polar angle,
the observed luminosity notably depends on the observer's viewing angle.
The supercritical accretion flow around the magnetized neutron star
tends to be identified as highly super-Eddington source since the 
radiative flux is enhanced around the magnetic axis.
In this study, we assume that the magnetic axis is parallel to the
rotation axis of the disks. Also, the non-rotating neutron star is
supposed. Thus, observed luminosity does not exhibit a periodic
modulation of the luminosity, which implies that our present model
cannot explain ULX pulsars. However, if the magnetic dipole axis
is misaligned with neutron star rotation axis and the neutron star
rotates, the observed luminosity
varies with the rotation period of the neutron star. If this is the
case, our results would support the hypothesis whereby
the ULX pulsars are powered by the supercritical accretion onto the
magnetized neutron stars.
Such argument would be confirmed by three-dimensional simulations.



We show that the net mass accretion rate is not constant, but it
increases for a larger radius ($r>8R_*$).
The non-steady accretion, especially non-uniform net mass accretion rate 
at larger radius, is caused by the shortage of computational time.
We initially set equilibrium torus at $r > 10R_*$. This gas torus
gradually falls onto the neutron star due to the development of MRI. 
Since the dynamical timescale is longer for larger radius, 
the system tends to become steady from the smaller radius. Thus, the
equilibrium radius, where the net mass inflow rate is constant, would
extend by long term simulations, although a steady state is not realized
for a larger radius $\gtrsim 8R_*$ in the present simulations. However,
such long term simulations are time-consuming and study of long term
evolution of disks is beyond of scope of this paper.

If such long term simulations are realized, the accretion rate might
increase due to the accretion of the remnant of the initial torus. Then,
the magnetosphere would shrink gradually. When the mass accretion rate
is sufficiently high, the radiation pressure of the disks overcomes the
magnetic pressure. Then the magnetosphere disappears and supercritical
disk accretion would be realized \citep{2007PASJ...59.1033O}.

It is noted that an effect of boundary condition would affect on results
for a long term simulations. We observed a gradual increase in mass
density on the neutron star surface. This is caused by the reflection
boundary condition by which the matter tends to accumulate around the
neutron star surface. Therefore, the mass density near the surface would
be larger in the long term simulations. If we adopt free boundary
condition, the mass is swallowed by the inner boundary and the increase
of the mass density would not occur. Such a boundary condition is
adopted by authors \citep[e.g.][]{2012MNRAS.421...63R}.

We adopted a truncation formalism for radiation field by adopting the
M-1 closure. The M-1 closure is useful to describe the radiation field
in optically thick limit and optically thin free-streaming limit. Also
since the computational cost is reasonably small, the M-1 closure is
adopted in many astrophysical fields. However, we caution that adopting
M-1 closure is sometimes not validated especially in optically thin,
anisotropic case. Since the moment formalism is obtained by integrating
transfer equation in solid angles, two radiation beams unphysically
collides each other \citep[e.g., for example][]{2016ApJ...818..162O}. A
similar situation might happen in our simulations, since the two
accretion columns (north and south) are main light sources. The photons
emitted by the north accretion columns can collide with that emitted by
the south accretion column around an equatorial plane. The photons propagate in
radial direction after collision. The radial stream lines of radiation
flux between two accretion columns might be caused by such collision
(see $r < 1.5R_*$ in the lower latitude in the right panel of figure
1-b). Such unphysical collision might affect on accretion dynamics. If
the photon collision does not occur, two accretion columns illuminate
each other. The radiation force might induce the time variability and/or
oscillation of accretion columns. However, it is difficult to conclude
whether radial profiles of stream line is a consequence of unphysical
merging of radiation flux. Even if the photons pass through without
collision, the $\theta$-component of the radiation flux becomes very small. Then, stream lines of radiation flux would be almost radial,
although two photons never merge.

Also the M-1 closure cannot treat radiative viscosity
correctly. Numerical studies showed that an efficiency of angular
momentum transport by MRI-driven turbulence depends on the magnetic
Prandtl number \citep{2007MNRAS.378.1471L,2007A&A...476.1123F,2009ApJ...707..833S}
\cite{2013ApJ...767..148J} performed MHD simulations with frequency
integrated radiation transfer equation and show that radiative viscosity
significantly exceeds microscopic viscosity and magnetic resistivity in
the context of local shearing box approximation. But they also showed
that it is still negligible compared to Maxwell stress and Reynolds
stress. On the other hand, we show that the angular momentum of the gas
inside accretion column is transported to the neutron star and is small
due to the magnetic torque by the dipole field. Thus the radiative
viscosity might not affect on the accretion dynamics both in accretion
column and accretion disks. Even though that, we have to solve transfer
equation for radiation to resolve these problems. \cite{2014ApJ...796..106J}
performed radiation magnetohydrodynamics simulations by solving
frequency-integrated transfer equations. They observed a lower radiative
efficiency than that of GR-RMHD results. But we need to pay attention
that they assume non-relativistic plasma. The discrepancy requires
further investigation. \cite{2016ApJ...818..162O} developed a numerical
method solving transfer equation in the framework of the special
relativity. Solving the transfer equation including the general
relativistic effect has not been succeeded yet and is the grand
challenge to understand accretion disks onto the black hole and neutron
star. We will tackle on this problem in a near future.

\begin{deluxetable}{lcccc}
\tablehead{
\colhead{} 
&\colhead{$\dot M_\mathrm{in}/M_\mathrm{crit}$} 
&\colhead{$\dot M_\mathrm{out}/M_\mathrm{crit}$} 
&\colhead{$\dot M_\mathrm{out}/M_\mathrm{in}$} 
&\colhead{$L_\mathrm{bol}/L_\mathrm{Edd}$} 
}
\startdata
$r=1.1R_*$ & 66   &  55 & 0.84 & 6.1 \\
$r=3R_*$   & 301  & 291 & 0.97 & 9.0 \\
$r=8R_*$   & 1350 & 935 & 0.69 & 7.0 \\
\enddata
\tablecomments{
From left to right, mass accretion rate, outflow rate, its ratio, and
 bolometric luminosity.
 Values measured at $r=1.1,\ 3,\ 8R_*,$ are time averaged between $t=10^4t_\mathrm{g}$
 and $t=1.5\times 10^4t_\mathrm{g}$.
}
\label{tab1}
\end{deluxetable}

\acknowledgments
Numerical computations were carried out on Cray XC30 at the Center for
Computational Astrophysics of National Astronomical Observatory of
Japan, on FX10 at Information Technology
Center of the University of Tokyo, and on K computer at AICS.
This work is supported in part by JSPS Grant-in-Aid for Young Scientists
(17K14260 H.R.T.) and for Scientific Research (C) (15K05036 K.O.) .
This research was also supported by MEXT as 'Priority Issue on Post-K
computer' (Elucidation of the Fundamental Laws and Evolution of the
Universe) and JICFuS.


\end{document}